\documentclass{elsart}
\usepackage[dvips]{graphicx}                  
\usepackage{amssymb}

\def\Tmelt{\ensuremath{T_\mathrm{m}}}

\def\gSV{\ensuremath{\gamma_\mathrm{SV}}}
\def\gLV{\ensuremath{\gamma_\mathrm{LV}}}
\def\gSL{\ensuremath{\gamma_\mathrm{SL}}}

\def\tLV{\ensuremath{\theta_\mathrm{LV}}}
\def\tSL{\ensuremath{\theta_\mathrm{SL}}}

\begin{document}
\begin{frontmatter}
\title{Physics and Nanofriction of Alkali Halide Solid Surfaces at the Melting Point}
\author[sissa,democritos]{T. Zykova-Timan\corauthref{corresponding}}
 \corauth[corresponding]{Corresponding author.}
   \ead{tzykova@sissa.it}
\author[sissa,democritos]{D. Ceresoli}
\author[sissa,democritos]{U. Tartaglino}
\author[sissa,democritos,ictp]{E. Tosatti}
 \address[sissa]{International School for Advanced Studies (SISSA), and
   INFM Democritos National Simulation Center, Via Beirut 2-4,
   I-34014 Trieste, Italy}
  \address[democritos]{INFM Democritos National Simulation Center,
   Trieste, Italy}
\address[ictp]{International Center for Theoretical Physics (ICTP),
   Strada Costiera 11, 34014, Trieste, Italy}

\begin{abstract}
Alkali halide (100) surfaces are anomalously poorly wetted by their
own melt at the triple point. We carried out simulations for NaCl(100) within 
a simple (BMHFT) model potential. Calculations of the solid-vapor, 
solid-liquid and liquid-vapor free energies showed that solid NaCl(100) 
is a nonmelting surface, and that the incomplete wetting can be traced  to the conspiracy
of three factors: surface anharmonicities stabilizing the solid
surface; a large density jump causing bad liquid-solid adhesion; incipient
NaCl molecular correlations destabilizing the liquid surface, reducing
in particular its entropy much below that of solid NaCl(100). 
Presently, we are making use of the nonmelting properties of this
surface to conduct case study simulations of hard tips sliding 
on a hot stable crystal surface. Preliminary results reveal novel phenomena
whose applicability is likely of greater generality.
\end{abstract}
\end{frontmatter}
\section{Introduction}
\begin{figure}[ht!]
    \begin{center}
    \includegraphics[width=0.6\textwidth]{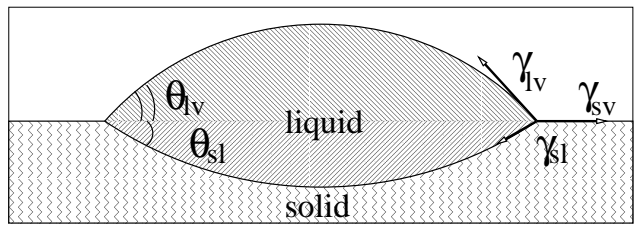}
    \label{scheme}\caption{The balance of the forces at the solid-liquid-vapor
    interface}.
    \end{center}
    \end{figure}

Most liquids in nature are expected to wet very well the surface of their own 
solid at the melting point; and in fact they generally do. However, that 
is not always true and some examples of nonwetting do exist~\cite{review}.
 A striking case in point is sodium chloride, where the NaCl(100) solid surface is 
badly wetted by its own melt, with an anomalously large partial contact 
angle of 48$^\circ$(Fig.~\ref{scheme})~\cite{mutaft}. How could one have predicted that, 
and what are the physical consequences? Moreover what applications, and what 
new phenomena, for instance in the field of friction, could this kind of 
non-self-wetting lead to? We addressed these questions using simple theory -- 
basically surface thermodynamics and applied statistical mechanics -- and 
computer simulation, in particular classical molecular dynamics (MD).
 \protect
    
Consider a NaCl droplet deposited on NaCl(100). From Fig.~\ref{scheme},
$\gSV = \gSL \cos~\tSL + \gLV \cos~\tLV$
and there will be partial wetting by the droplet if 
the three interface free energies satisfy $\gSV < \gSL
+ \gLV$ ($\tSL \simeq 0$ in our case). To obtain a microscopic theory, 
we must be able to calculate these three interface free energies. 
For the liquid surface, 
there is a well established method, the Kirkwood-Buff virial formula~\cite{PRL,long}, that builds the 
surface stress (surface tension), equal to the interface free energy, 
from mechanical variables, {\it i.e.,} from forces and coordinates that can in 
turn be extracted from MD simulations.
   Nothing so straightforward is available for the solid-vapor 
and for the solid-liquid interfaces. The problem is that the solid 
can support stress, so that the surface stress, which is mechanical and 
easily calculable, and the surface free energy are no longer the same. 
For the solid-vapor interface, 
the problem is solved by resorting to two parallel thermodynamic integrations
carried out for two systems, identical in every respect, {\it i.e.,} same number 
of particles, same crystal structure, same cell etc., except that one system is 
a bulk without surfaces, the other is a slab with two (identical) surfaces. 
Integration of $  \int_{1/T_i}^{1/T} E(T')\,
d\left(\frac{1}{T'}\right)$ from some low temperature $T_i$ up to 
$T$ gives the free energies of the two systems. The surface free energy 
is obtained at each $T$ as half their difference (the slab has two surfaces). 
For the solid-liquid 
interface finally, not even that works, because the interface is stable just at 
$T$ = $\Tmelt$. Here, we circumvented the difficulty by turning so to speak 
experimentalists. We simulated a NaCl nanodroplet onto NaCl(100), let it 
settle, and extracted the contact angle $\tLV$~\cite{droplet}. Once $\tLV$, 
$\gSV$ and $\gLV$ are known, then $\gSL$ can easily be obtained from Young's 
force balance equation given above.
\section{ Calculations and Results}
All results were  produced using classical MD simulations, mostly in slab 
geometries. Na$^{+}$ and Cl$^{-}$ ions were assumed to interact via BMHFT 
potentials~\cite{fumi}. These potentials are standard 
and well tested, but they have a long range Coulomb part. 
This reduces dramatically the sizes of systems we can simulate -- we used 
between 1500 and 5000 NaCl molecular units -- and the simulation times we 
could afford. Luckily, two hundred psec were usually more than enough to attain 
equilibrium configurations at our high temperatures. We ran a bulk simulation 
first, and extracted the bulk melting $\Tmelt$ -- the temperature where the BMHFT
solid and liquid coexist -- at 1066 K, extremely close to the experimental 
1074\,K of NaCl~\cite{PRL,long}.

\begin{figure}
    \begin{center}
    \includegraphics[width=0.8\textwidth]{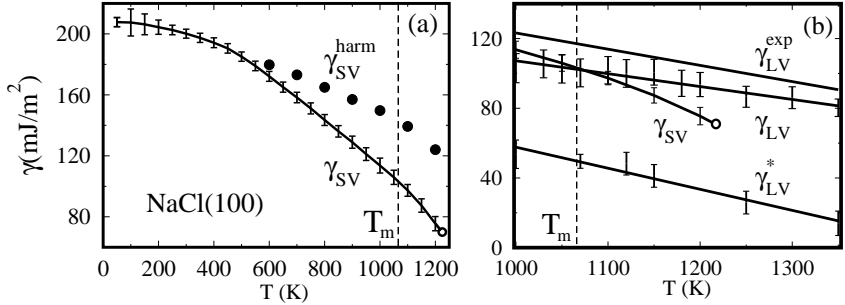} \\
    \caption{Calculated NaCl surface free energies. a) Solid-vapor
    $\gSV$.  Note the metastability up to about 150~K above
    $\Tmelt$.  Dots: effective harmonic approximation. b) Liquid-vapor
    free energy: experimental and simulated values.\label{gammas}} 
    \end{center}
    \end{figure}

We then simulated solid NaCl(100), and found that it remained 
dry and crystalline up to 1210 K, well above the bulk $\Tmelt$.
This kind of nonmelting behavior, where the solid surface can survive in a
metastable state even above $\Tmelt$, is exactly what can be expected when
the liquid does not wet the solid~\cite{review}. 
The same slab system was then melted 
to generate a liquid slab -- a liquid with two liquid-vapor interfaces. 
From these simulations, we extracted as explained earlier $\gSV$
and  $\gLV$ shown in Fig.\ref{gammas}. Here came three surprises. 
The first was the large decrease of $\gSV$ with increasing $T$, strongly stabilizing the 
solid surface at high temperature. This is in part an anharmonic effect, 
and stems from the exceptional stability of long range Coulomb systems against 
large vibration amplitudes -- even in bulk alkali halides largely exceeding the Lindemann 
melting criterion~\cite{long}. The second surprise was to find  $\gLV$ so high -- 
numerically equal to  $\gSV$ at $\Tmelt$, when usually the solid surface 
is energetically much more expensive than the liquid one. The third 
surprise  finally was that the 
{\it surface entropy} $-d\gamma/dT$ is a factor nearly three {\it lower}
in the liquid than in the solid. This is strikingly contrary to our perception 
of a very disordered liquid surface (Fig.3) as opposed to an ordered solid surface. 
As it turns out these two last surprises are related to one another, and are 
related to incipient {\it molecular} NaCl pair bond formation in the outer 
layer of the liquid surface. A way to assess quantitatively this effect was
found by recalculating the surface tension
after removing molecular correlations, and showing that when this is done both
$\gLV$ (see $\gLV^{\star}$ in Fig.\ref{gammas}(b)) and the liquid surface 
entropy drop hugely~\cite{PRL,long}.
    \begin{figure}\begin{center}
    \includegraphics[width=0.5\textwidth]{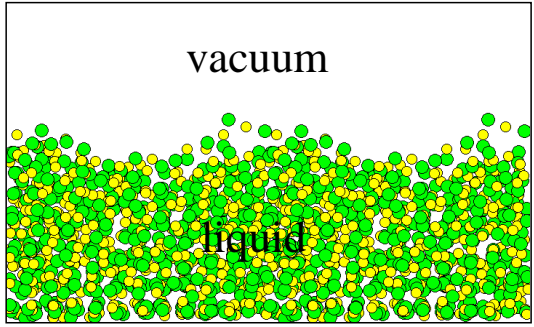}
    \caption{Snapshot of simulated liquid-vapor interface
    at $\Tmelt$\label{liquid}}. 
    \end{center}\end{figure}
    
\begin{figure}\begin{center}
    \includegraphics[width=0.8\textwidth]{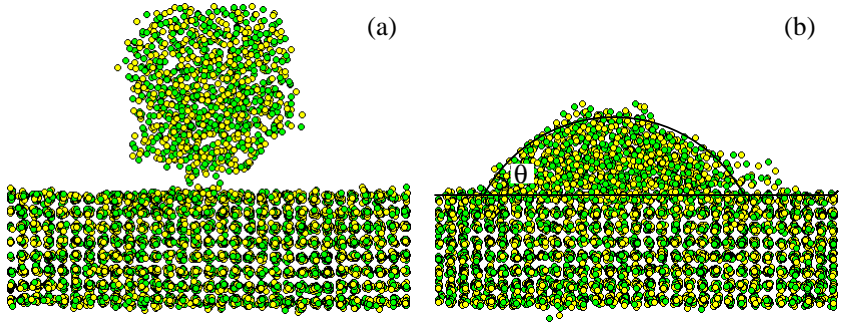}
    \caption{Initial (a) and final (b) snapshots of NaCl droplet deposition on
    NaCl(100) at $T$ = $\Tmelt$. For the last 100 ps the drop remains in a metastable 
    state forming a negligible internal contact angle $\tSL$, and an external contact $\tLV$ 
    angle of 50$^\circ$.\label{drop}} 
    \end{center}\end{figure}

Finally, we melted a small solid NaCl cube to form a nanodroplet, 
and deposited it on the solid NaCl(100) slab surface (Fig.~\ref{drop}).
From this final simulation we extracted an external wetting angle $\tLV$
of 50$\pm$5$^\circ$, in good agreement with the experimental 48$^\circ$. 
When inserted in the force balance equation along with $\tSL \sim 0 $,
that finally yields $\gSL$ = 36 mJ/m$^{2}$, which is about 1/3 of  $\gSV$ 
and $\gLV$. This relatively large value clearly demonstrates the poor adhesion 
of liquid NaCl to the solid, in turn reflecting the fact that they are very 
different -- for example the liquid is 27\% less dense than the solid.
Nucleation studies~\cite{expt,frenkel2005} suggested even larger values of $\gSL$
as large as 80 mJ/m$^{2}$ at $T \sim$800\,K.

In conclusion, three separate pieces of physics conspire to cause the
poor wetting properties of NaCl(100) by its own melt. The first is
the extreme anharmonic stability of the solid surface itself. The second
is the poor adhesion of the liquid to the solid. The third is the high liquid 
surface tension caused by the unexpected but very real surface entropy 
deficit~\cite{PRL,long}.
               
\section{Nanofriction}
The next class of problems we are presently addressing, 
is as important as it is unexplored, is the high temperature nanofriction of 
a hard sliding tip near $\Tmelt$ for a nonmelting surface like NaCl(100). 
On a normal solid surface -- one that is wetted by its own liquid -- a thin liquid film 
will nucleate below $\Tmelt$ causing a jump to contact with the approaching tip~\cite{frenken},
and generally ruining it. A nonwetting surface like NaCl(100) will not do such thing.
It is thus ideal, at least theoretically, to explore friction close to the melting point.
Friction has been the subject of important simulation work~\cite{persson,landman}.
High temperature friction is only recently beginning to receive some attention~\cite{thermolub}
and there is a clear need for more work.We intend precisely to explore and motivate future
work in this extreme regime. 

\begin{figure}\begin{center}
    \includegraphics[width=0.4\textwidth]{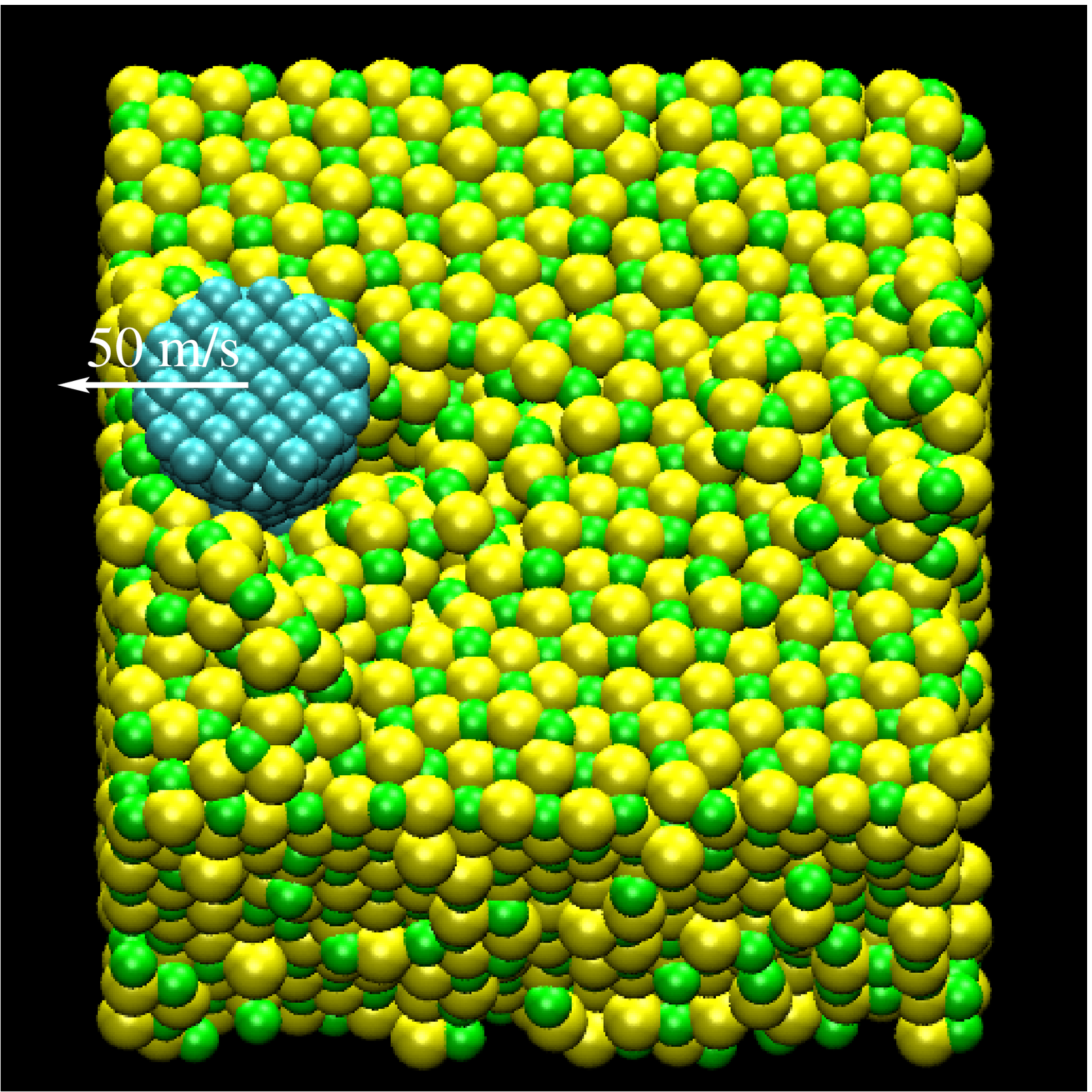}
    \hfill\includegraphics[width=0.4\textwidth]{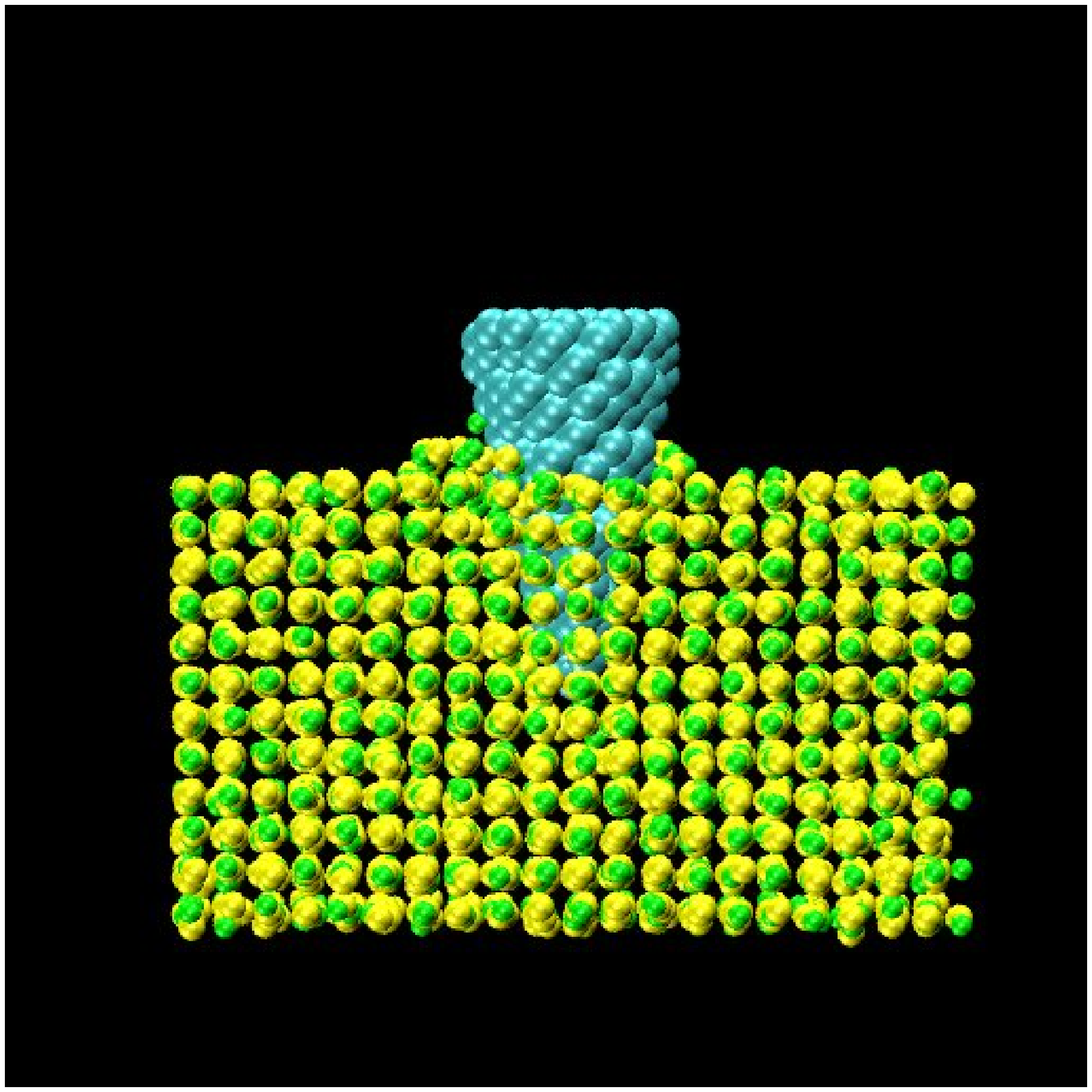}\\
     \caption{Nanofriction simulation of a hard diamond tip ploughing into NaCl(100)
     near $\Tmelt$. The substrate is solid, but despite that the furrow heals away and
     recrystallizes in a very short time of a few ps.\label{frict}}   
    \end{center}\end{figure}

\begin{figure}\begin{center}
    \includegraphics[width=0.5\textwidth]{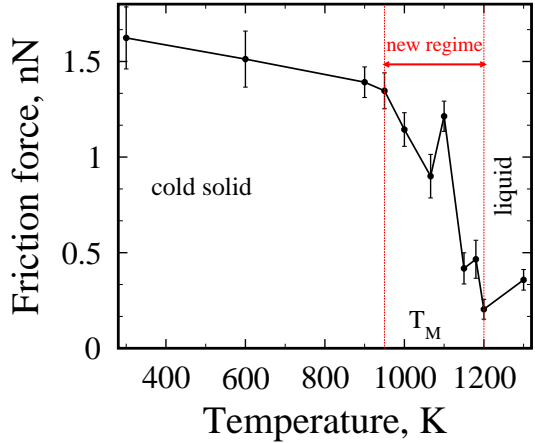}
    \caption{The dependence of friction force on temperature
     shows a decrease of nanofriction on NaCl(100) close to the melting point,
     as in ice skating.\label{skate}}
    \end{center}\end{figure}

   We conducted preliminary sliding friction simulations for hard tips on NaCl(100),
exploring both the deep ploughing regime with a sharp tip (Fig.~\ref{frict}), and the gentle
grazing regime with a flat tip. The tip was modeled to reproduce the
interaction between rigid diamond and the NaCl surface\cite{tang,nanof}. For deep wear
simulations  we created a conical tip,
composed of $\sim$ 400 atoms. The diameter was around 13\AA\ and the height $\sim$ 26\AA.
On the contrary for light grazing
wearless friction we use a completely blunt flat tip composed of
$\sim$ 200 atoms (not shown). The flat surface of the
tip has a (111) plane of 13\AA\ diameter roughly. 
In ploughing regime the tip was moving with a constant speed of 50 m/sec and the constant
penetration depth of 6\AA.
The wear 
frictional force of Fig.\ref{skate} shows that
at high temperature there is a regime where friction drops, although the
surface is still fully solid. The tip scratches the solid surface, leaving 
behind a furrow which in this regime heals out closing spontaneously  
on a very short time of 10 ps. Direct inspection shows that the tip is surrounded in this 
high temperature regime by a seemingly liquid cloud, moving along with it and
lubricating its motion. In this way, the tip effectively {\it skates} over the 
solid NaCl(100) surface. Discussions of skating near the melting point appear to exist
so far only for ice~\cite{persson}.

Preliminary results show that under grazing conditions, the high temperature behavior of 
sliding friction is just the opposite~\cite{nanof}. The initially very low sliding friction 
of the flat tip over the cold surface rises with temperature, culminating
in a frictional peak very close to the melting point. Here it appears that 
the large compliance of the (nearly unstable) hot NaCl surface lattice
is responsible for the increased high temperature friction. There is to our knowledge
no prior established mechanism of this kind for high temperature
frictional increase.

These predicted nanofriction phenomena are currently under closer scrutiny
for further physical characterization. It is hoped that experimental efforts
could be started to investigate their existence, which we believe to be more
general for all nonmelting surfaces, rather than specific to the NaCl(100) 
system studied here. A more detailed report of this part of the work is 
forthcoming~\cite{nanof}.

\section*{Acknowledgements}
This work was partly supported by MIUR COFIN No. 2003028141-007, 
MIUR COFIN No. 2004023199-003, by  FIRB RBAU017S8R operated by INFM,
by MIUR FIRB RBAU017S8, and by INFM (Iniziativa trasversale calcolo parallelo).



\end{document}